\documentclass[10pt, twocolumn]{article}
\usepackage{latexsym,graphicx,multirow}
\usepackage{amssymb}
\usepackage{amsmath}
\usepackage{amscd}
\usepackage{amsthm}
\usepackage[left=2cm,top=2.5cm,right=2.5cm,bottom=1.5cm]{geometry}
\usepackage{hyperref}
\usepackage{color}

\usepackage{epstopdf}
\usepackage{cite}
\usepackage{float}
\usepackage[utf8]{inputenc}

\begin{document}
\twocolumn[
  \begin{@twocolumnfalse}
    \begin{center}
        \large{\bf{New Tsallis Agegraphic Dark Energy}} \\
        \vspace{10mm}
   \normalsize{ Pankaj$^{a,1}$, Bramha Dutta Pandey$^{b,2}$, P. Suresh Kumar$^{c,3}$, Umesh Kumar Sharma$^{d,4}$ } \\
    \vspace{5mm}
    \normalsize{$^{a}$ IT Department (Math Section), University of Technology and Applied Sciences-HCT, Muscat, Oman }\\
    \vspace{2mm}
    \normalsize{$^{b,\:c}$ IT Department (Math Section), University of Technology and Applied Sciences-Salalah, Oman}\\
    \vspace{2mm}
    \normalsize{$^{b,\:c,\:d}$ Department of Mathematics, Institute of Applied Sciences and Humanities, GLA University
    	Mathura-281406, Uttar Pradesh, India}\\
    \vspace{2mm}

    $^1$E-mail: pankaj.fellow@yahoo.co.in \\
    $^2$E-mail: bdpandey05@gmail.com \\
    $^3$E-mail: sureshharinirahav@gmail.com \\
    $^4$E-mail: sharma.umesh@gla.ac.in \\
    \vspace{10mm}

\end{center}

\begin{abstract}
The proposed model is a study of the nature of dark energy through non-extensive Tsallis entropy. The method is based on the Karolyhazy relation which is a combined idea from quantum physics and general relativity. Dark energy is the energy density of quantum fluctuations in space-time. This is the key idea behind proposing agegraphic dark energy (ADE) models here. The parameter $\delta$ is used to measure the quantitative distinction from the standard scenario. To look at the cosmological implications of the hypothesized dark energy model, as well as the expansion of the Universe filled with zero pressure matter and the resulting dark energy alternatives, the role of IR cutoff is played by age of the universe. The dynamic behavior of dark energy density parameter is carried out. The expressions for the equation of state parameter and deceleration parameter are obtained. The analysis is performed by taking into account a no flow  as well as  a flow of energy among the dark matter and dark energy sectors of the universe. \\
\smallskip

{\bf Keywords}: NTADE, Dark energy, Tsallis entropy, Cosmology \\
PACS:  95.36.+x, 98.80.Es, 98.80.Ck\\

\end{abstract}

\end{@twocolumnfalse}]

\section{\textbf{Introduction}}
Since the discovery of accelerated expansion of our universe  \cite{Riess98,Perl99,Planck:2018vyg}, dark energy (DE) has been one of the most active disciplines in modern cosmology. DE is the mysterious energy which  is responsible for the Universe's current accelerated expansion phase and accounts for roughly $70\%$ of the universe's energy content \cite{Peebles:2002gy, Bamba:2012cp, Vagnozzi:2021quy,Goswami:2019zto}. The cosmological constant, which is a major constituent of the $\Lambda-$CDM model, with $P_d=-\rho_d$ is the simplest choice for the DE \cite{Padmanabhan02,Sahni99,Copeland:2006wr}. But the $\Lambda-$CDM model shows inability in dealing with fine-tuning and cosmic coincidence issues \cite{Padmanabhan02,Peebles:2002gy}. Such circumstances force researchers to look for alternative DE models.

The cosmological constant can be measured through gravitational tests, being related to some quantum field's expectation measure. And hence it becomes a problem of quantum gravity. A mixed approach of General relativity and quantum gravity may be an approach to solve the said issue. As per general relativity theory, laws of physics related to space-time can be verified to any desired accuracy. But quantum physics puts the accuracy bound, thanks to Heisenberg's uncertainty relation. Using the principles of quantum physics and general relativity, Karolyhazy and his partners \cite{Karolyhazy66} made an important finding on measuring the distance for Minkowski space-time, inline with quantum fluctuations of space-time \cite{Maziashvili06}. Karolyhazy with his colleagues uncovered an interesting fact regarding measuring distance in Minkowski spacetime based on the quantum fluctuations of spacetime \cite{Karolyhazy66, Maziashvili06}. According to which the distance $t$ in Minkowski spacetime cannot have accuracy better than $\delta t=\lambda t_p^{\frac{2}{3}}t^{\frac{1}{3}}$ with dimensionless constant $\lambda$ and reduced Plank time $t_p$. Based on  Karolyhazy relation and as an alternative to  $\Lambda$ - CDM model, Cai \cite{Cai07} and Wei and Cai \cite{Wei08}  introduced the agegraphic dark energy (ADE)model by considering age of the universe and conformal time as an infrared (IR) cutoff. Recent results show the dependent evolution of DE and dark matter (DM) sectors of the universe \cite{Amendola99,Zimdahl01, Chimento03, delCampo06, Wang16, Pavon05, DiValentino:2019ffd, Gomez-Valent:2020mqn} that can solve the  "problem of coincidence" \cite{Salvatelli14}. In \cite{Wei09}, the interacting ADE model is studied using the power law of entropy and age of the universe as IR cutoff. The authors studied the model for interacting and non interacting sectors of the cosmos and obtained the expressions for ADE density, EoS and deceleration parameters in both the situations \cite{Wei07, Cui09, Fazlollahi:2022dwz,Kim08,Sheykhi09a,Sheykhi09b, Zhang:2008mb, Wei:2007zs, Jamil:2010vr, Setare:2010zy, Li:2010ak}.

Motivated by Gibbs \cite{Gibbs1902}, Tsallis \cite{Tsallis88} stated the statistical description for non-extensive systems which gives a new entropy-area relationship for horizons \cite{Tsallis12}. The Tsallis principle states that to each black hole, there is an associated entropy. This entropy is described as $S_\delta=\gamma A^\delta$, where $\delta$ is a non-extensive parameter and $\gamma$ is a constant. Recently\cite{Abdollahi18} and \cite{Sharma20} studied the ADE models in the line of \cite{Cai07, Wei08} using Tsallis and Barrow entropy respectively for interaction as well as non-interaction among the cosmos sectors. Several authors obtained various cosmic parameter expressions, stability, statefinder and phase space analysis for both interacting and non-interacting Tsallis agegraphic dark energy (TADE)  models and also studied TADE models in modified theory of gravity \cite{Srivastava:2021apm, Huang:2021rpf, FeiziMangoudehi:2022yvu, Kumar:2022acs, Sharma:2020mzl, Sardar:2021eaj, Srivastava:2020riu, Xu:2019hhs, Maity:2019qbv, Pradhan:2021tij}.

Motivated with above literatures, in this paper, we have proposed the new Tsallis ADE (NTADE) model with universe age and conformal time as an IR cutoffs with and without interaction. The manuscript will split as follows: Section 2 will be devoted to the formulation of the Tsallis agegraphic dark energy model. In Section 3, the results will be obtained by considering the two sectors to evolve independently. One full section will be dedicated for the interlink among the DM and DE sectors of the universe. Finally, section 5 will result in summarizing the  obtained results discussions. 

\section{Formulation of New Tsallis Agegraphic Dark Energy (NTADE) Model}
The DE density $\rho_d$ of a black with radius $L$ is related by $\rho_d L^4\leq S$ \cite{Li04}. This inequality is saturated as $\rho_d=\dfrac{3c^2M_p^2}{L^2}$ with model parameter $c$. Hence, modifying the entropy of a system results in a modified HDE model. Following Pandey et al. \cite{Pandey21}, the new Tsallis holographic dark energy density $\rho_d$ is defined by 
\begin{equation}
	\rho_d=\dfrac{3c^2M_p^2}{L^2}+\delta M_p^4+\delta^2M_p^6L^2.  \label{3.1}
\end{equation}
As the universe can't be older than its constituents, so the age of the universe $T$ will be considered as the IR cutoff. And hence, the equation (\ref{3.1}) leads to the new Tsallis agegraphic dark energy model whose energy density is given by
\begin{equation}
	\rho_d=\dfrac{3c^2M_p^2}{T^2}+\delta M_p^4+\delta^2M_p^6T^2, \label{3.2}
\end{equation} 
where $T$ is defined by
\begin{equation}
	T=\int_0^{a(t)} \mathrm{d}t=\int_0^{a(t)}\dfrac{\mathrm{d}a(t)}{Ha(t)}. \label{3.3}
\end{equation} 
On considering the Friedmann-Robertson-Walker (FRW) model's geometry is homogeneous, isotropic, and flat, with a metric containing the scale factor $a(t)$ defined by  
\begin{equation}
	\mathrm{d}s^2=-\mathrm{d}t^2+(a(t))^2\delta_{ij}\mathrm{d}x^i\mathrm{d}x^j, \label{3.3a}
\end{equation}
The scale factor $a(t)$ and the Hubble parameter $H$ are connected by $H=\frac{\dot{a}(t)}{a(t)}$.
The conservation law of energy and momentum  with interaction term $Q$ is given by 

\begin{equation}
	\dot{\rho_m}+3H\rho_m=Q, \label{3.4}
\end{equation}
\begin{equation}
	\dot{\rho_d}+3H\rho_d \left(1+w_d\right)=-Q,  \label{3.5}
\end{equation}
where $\rho_m$ represents the DM density. $w_d$ in equation (\ref{3.5}), is the EoS parameter defined by $w_d=\dfrac{P_d}{\rho_d}$ with DE pressure $P_d$. The positive value of $Q>0$ corresponds to flow of energy from DM to DE sector whereas negative $Q$ corresponds to DE to DM energy flow \cite{Wang16}.
In a flat FRW universe filled with a pressure-less fluid (density $\rho_m$) and TADE (density $\rho_d$), the first Friedmann equation is expressed as 
\begin{equation}
	3M_p^2H^2=\rho_d\;+\;\rho_m,    \label{3.6}
\end{equation}
which can be rewritten as 
\begin{equation}
	\Omega_d+\Omega_m=1,   \label{3.7}
\end{equation}
with NTADE and DM density parameters 
\begin{eqnarray}
	\Omega_m &=& \dfrac{\rho_m}{3M_p^2H^2}   \notag \\
	\Omega_d &=& \dfrac{\rho_m}{3M_p^2H^2},   \label{3.7a} 
\end{eqnarray}
respectively.

\section{Non Interacting NTADE model} 
In this section we will consider that the evolution of DE and DM sectors evolve independently, i.e. $Q=0$. Differentiating (\ref{3.2}) w.r.t. time and using equation (\ref{3.5}), we get the expression for the EoS parameter given by 
\begin{eqnarray}
	w_d &&=-1 \notag \\
	&& -\dfrac{2}{3HT}\left(\frac{\delta^2M_p^4T^4-3c^2}{\delta^2M_p^4T^4+\delta M_p^2T^2+3c^2}\right), \label{3.8}
\end{eqnarray}

where $T$ is given by  
\begin{flalign} 
	& T =\notag \\ 
	& \sqrt{\frac{\left(3H^2\Omega_d-\delta M_p^2\right)-\sqrt{\left(\delta M_p^2-3H^2\Omega_d\right)^2-12M_p^4c^2\delta^2}}{2\delta^2M_p^4}}.   \label{3.8a}
\end{flalign}

Differentiating (\ref{3.6}) w.r.t. time, using (\ref{3.5})and (\ref{3.7}) we get 
\begin{eqnarray}
	\dfrac{\dot{H}}{H^2} &&=-\dfrac{3}{2}(1-\Omega_d)\notag \\ &&+\frac{\Omega_d}{HT}\left(\dfrac{\delta^2M_p^4T^4-3c^2}{\delta^2M_p^4T^4+\delta M_p^2T^2+3c^2}\right). \label{3.9}
\end{eqnarray}

Using (\ref{3.9}), the expression for the deceleration parameter given by

\begin{eqnarray}
	&& q =-1-\dfrac{\dot{H}}{H^2}
	=\dfrac{1}{2}(1-3\Omega_d) \notag \\
	&&-\dfrac{\Omega_d}{HT}\left(\dfrac{\delta^2M_p^4T^4-3c^2}{\delta^2M_p^4T^4+\delta M_p^2T^2+3c^2}\right). \label{3.10}
\end{eqnarray}

Taking time derivative of (\ref{3.7a}) and using (\ref{3.5}), the differential equation for the NTADE density parameter is obtained as
\begin{eqnarray}
	&& \dot{\Omega_d}=2H(1+q)\Omega_d \notag
	\\ && +\dfrac{2\Omega_d}{T}\left(\dfrac{\delta^2M_p^4T^4-3c^2}{\delta^2M_p^4T^4+\delta M_p^2T^2+3c^2}\right). \label{3.11}
\end{eqnarray}

\begin{figure}[H]
	\centering
	\includegraphics[width=0.8\linewidth]{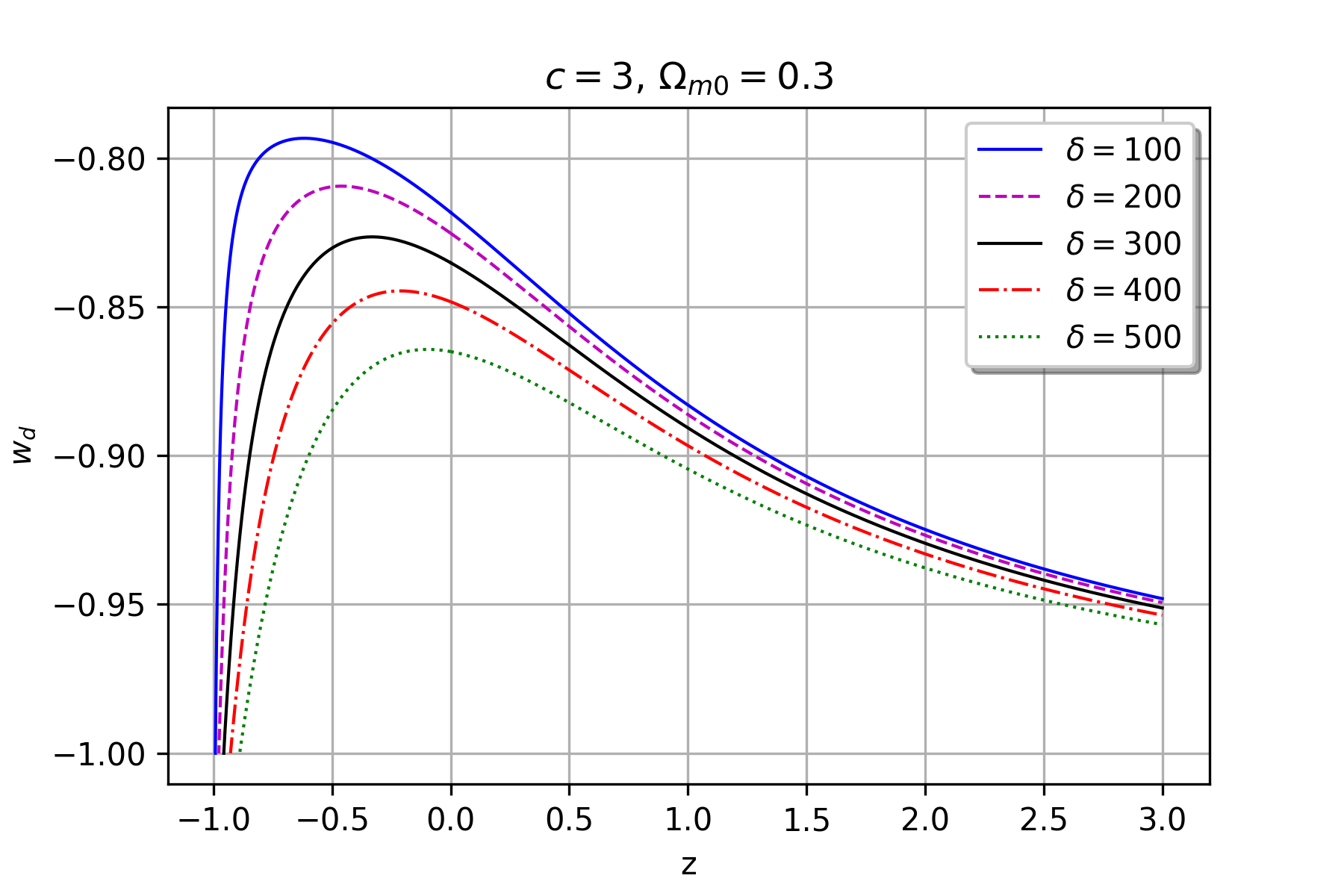}
	
	\caption{\begin{small}
			Behavior of $w_d$ vs $z$ for $H(z=0)=67.8, \quad M_p^2=1$ and varying values of $\delta$.
	\end{small}}
	\label{P3.1} 
\end{figure}

\begin{figure}[H]
	\centering
	\includegraphics[width=7cm,height=5cm]{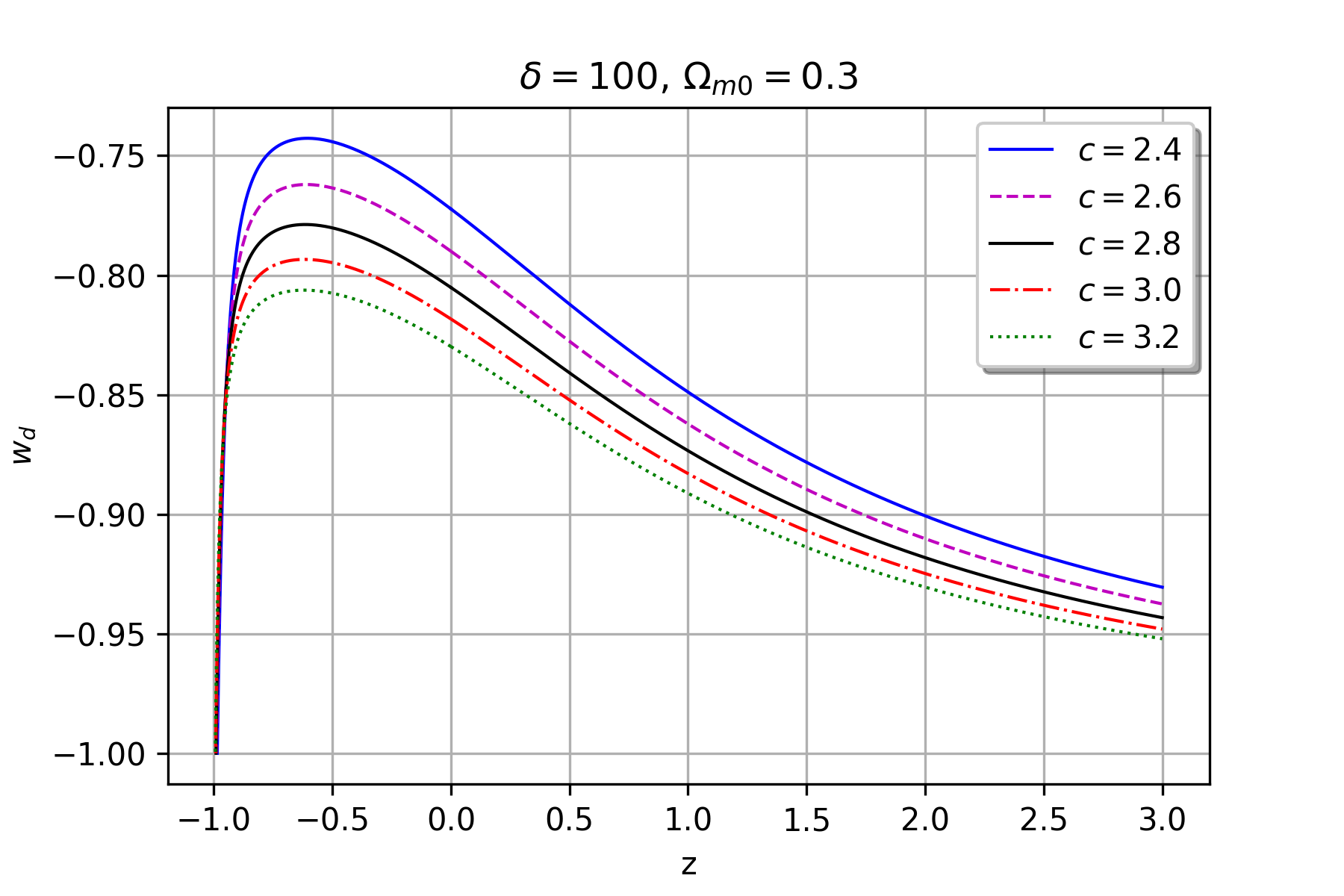}
	
	\caption{\begin{small}
			Behavior of $w_d$ vs $z$ for $H(z=0)=67.8, \quad M_p^2=1$ and varying  $c$ values.
	\end{small}}
	\label{P3.2} 
\end{figure}

Figures \ref{P3.1} and \ref{P3.2} show the EoS parameter behavior for the non interacting NTADE model. In plotting figure \ref{P3.1}, the parameter $c$ is fixed at $3$ and $\delta$ is allowed to vary from 100 to 500. For figure \ref{P3.2}, $c$ is allowed to vary by fixing $\delta$ at $100$. All the curves in the above two plots evolve in the quintessence region and finally, in the near or far future, meet the line $w_d=-1$. And hence, both the figures indicate that the non-interacting NTADE model is a pure quintessence model.
\vspace{1pt}
\begin{figure}[H]
	\centering
	\includegraphics[width=0.8\linewidth]{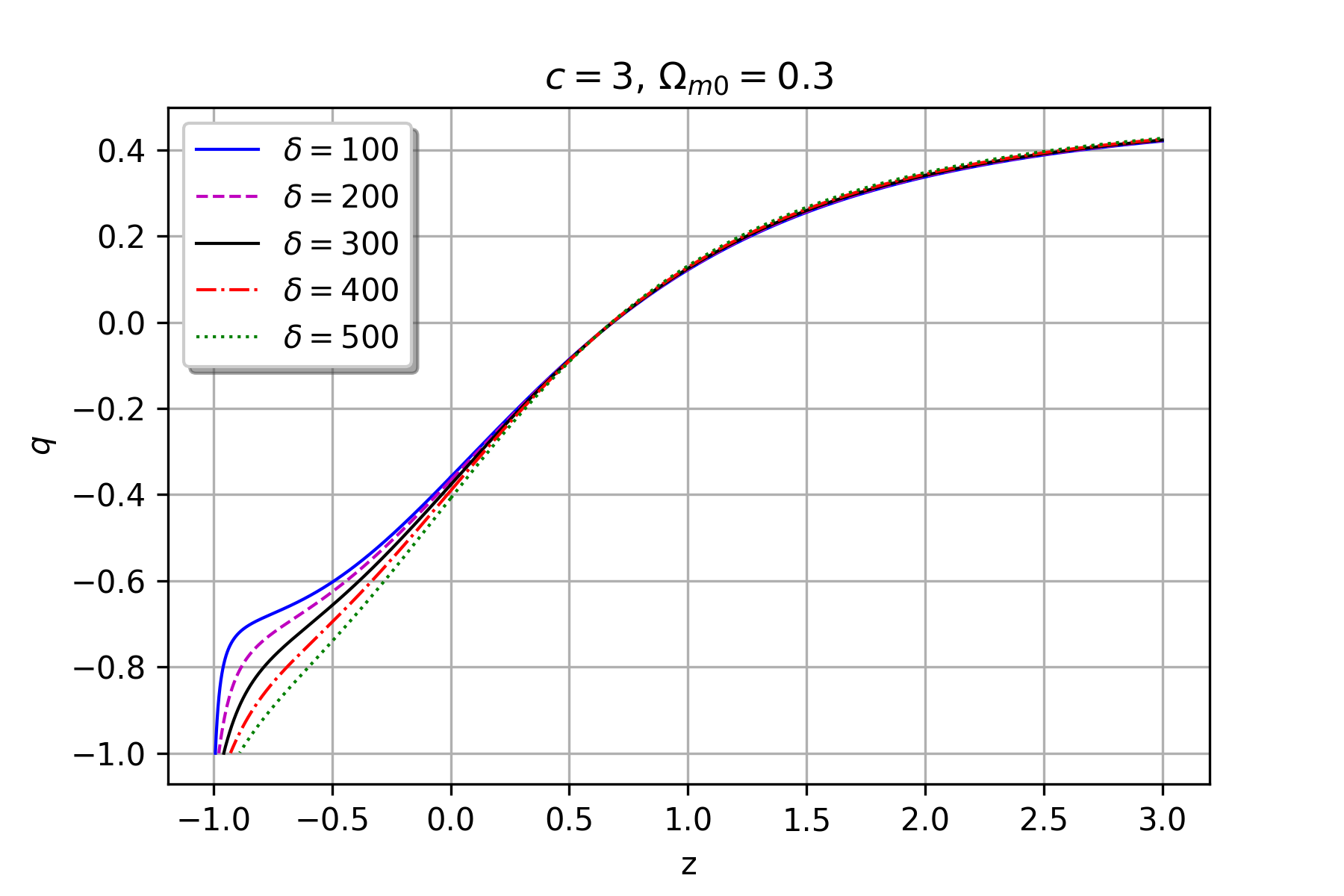}
	\caption{\begin{small}
			Behavior of $q$ vs $z$ for $H(z=0)=67.8, \quad M_p^2=1$ and varying values of $\delta$.
	\end{small}}
	\label{P3.3} 
\end{figure}
\vspace{1pt}
\begin{figure}[H]
	\centering
	\includegraphics[width=0.8\linewidth]{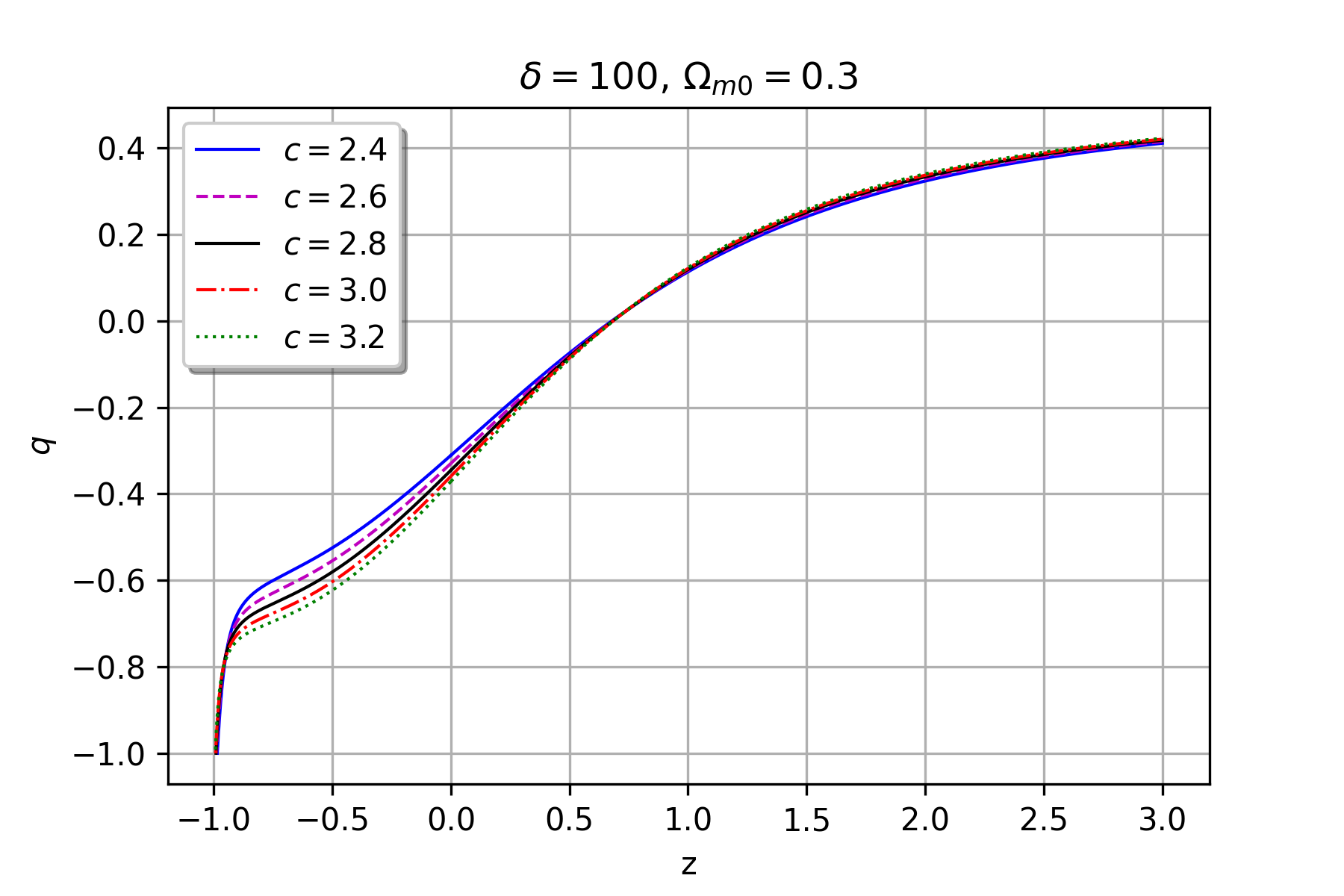}
	\caption{\begin{small}
			Behavior of $q$ vs $z$ for $H(z=0)=67.8,  \quad M_p^2=1$ and prescribed $c$ values.
	\end{small}}
	\label{P3.4} 
\end{figure}
\vspace{1pt}
The deceleration parameter $q$ is plotted in figures \ref{P3.3}-\ref{P3.4}. In plotting figure \ref{P3.3}, $\delta$ is varying with $c=3$ whereas in figure \ref{P3.4}, $c$ is allowed to vary by keeping $\delta$ at $100$. Curves in figure \ref{P3.3} and \ref{P3.4} indicate that the universe evolved from a decelerated phase. As $z$ approaches $0.6$ from right, the deceleration comes to an end. The acceleration started after $z\approx 0.6$.Clearly, the current universe is under accelerated expansion.

\begin{figure}[H]
	\centering
	\includegraphics[width=0.8\linewidth]{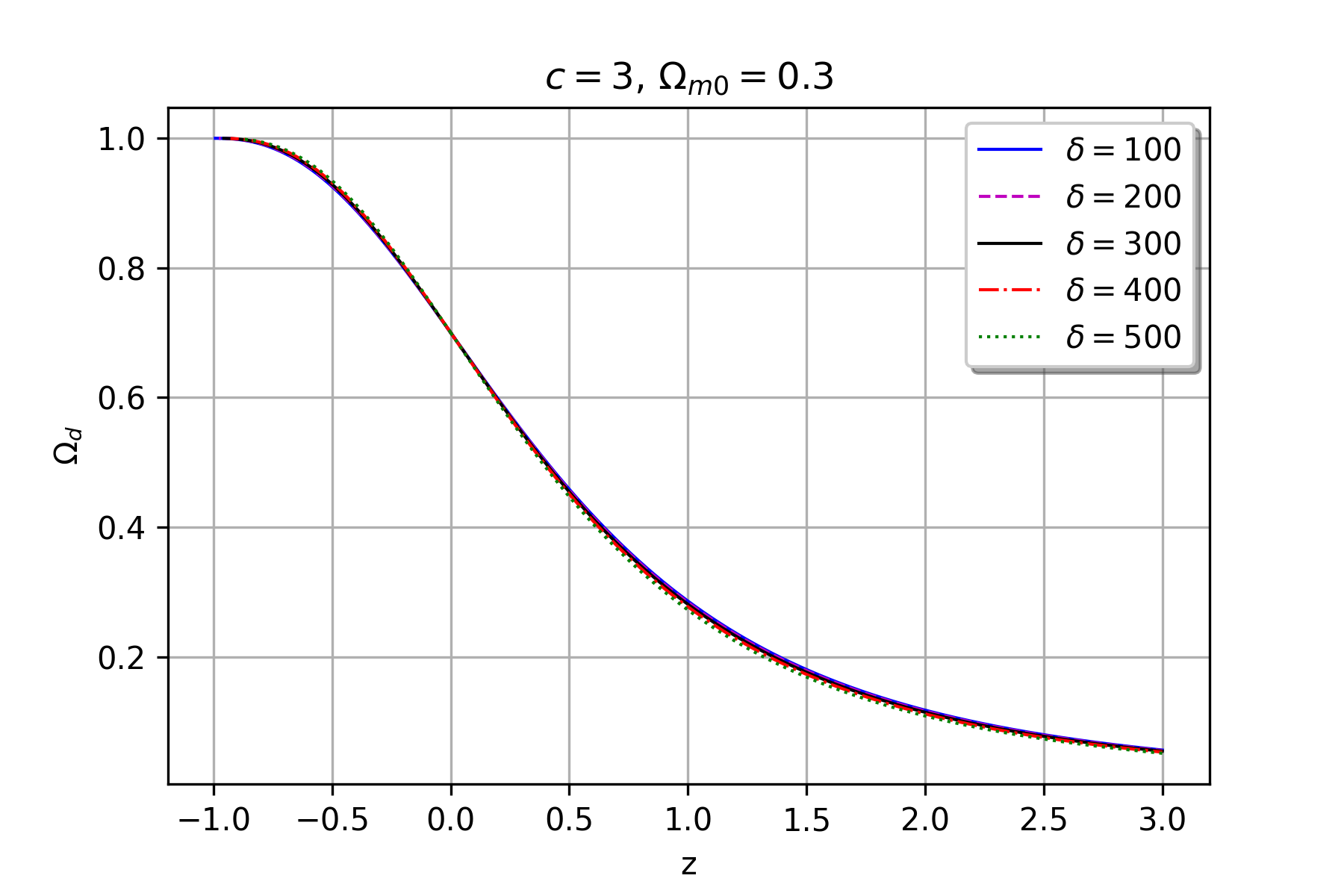}
	
	\caption{\begin{small}
			Behavior of $\Omega_d$ vs $z$ for $H(z=0)=67.8,  \quad M_p^2=1$ and varying values of $\delta$.
	\end{small}}
	\label{P3.5} 
\end{figure}
\vspace{0pt}
\begin{figure}[H]
	\centering
	\includegraphics[width=0.8\linewidth]{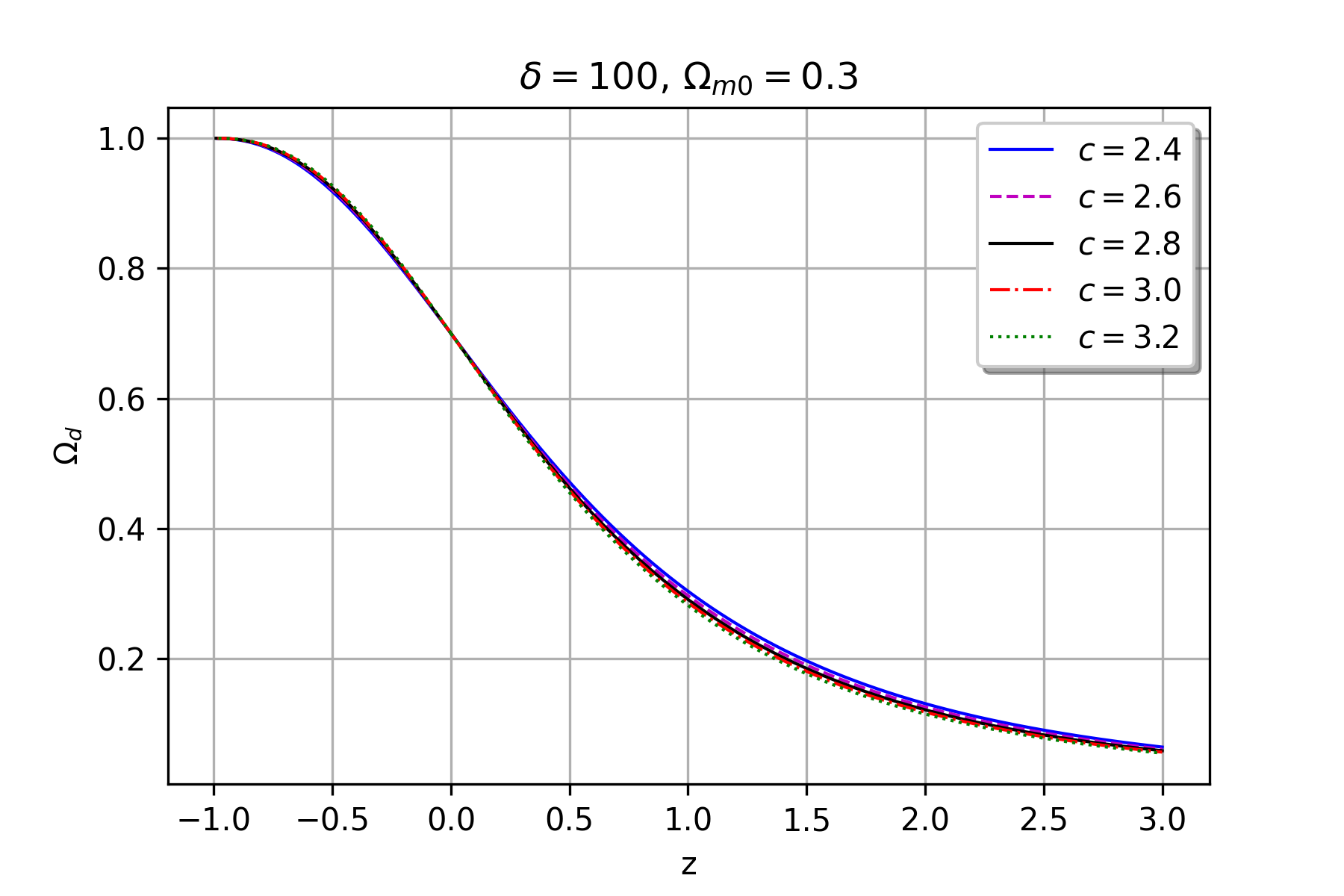}
	
	\caption{\begin{small}
			Behavior of $\Omega_d$ vs $z$ for $H(z=0)=67.8, \quad M_p^2=1$ and prescribed $c$ values.
	\end{small}}
	\label{P3.6} 
\end{figure}
\vspace{0pt}
Figures \ref{P3.5} and \ref{P3.6} are plotted for the NTADE density parameter $\Omega_d$ by considering variation in $\delta$ and in $c$, respectively.Both the curves show that the universe was dominated by DM in the past. Somewhere in $z\in (0.4, 0.6)$, the NTADE started dominating and indicated to overtake fully in the near or far future.
\vspace{0pt}
\begin{figure}[H]
	\centering
	\includegraphics[width=0.8\linewidth]{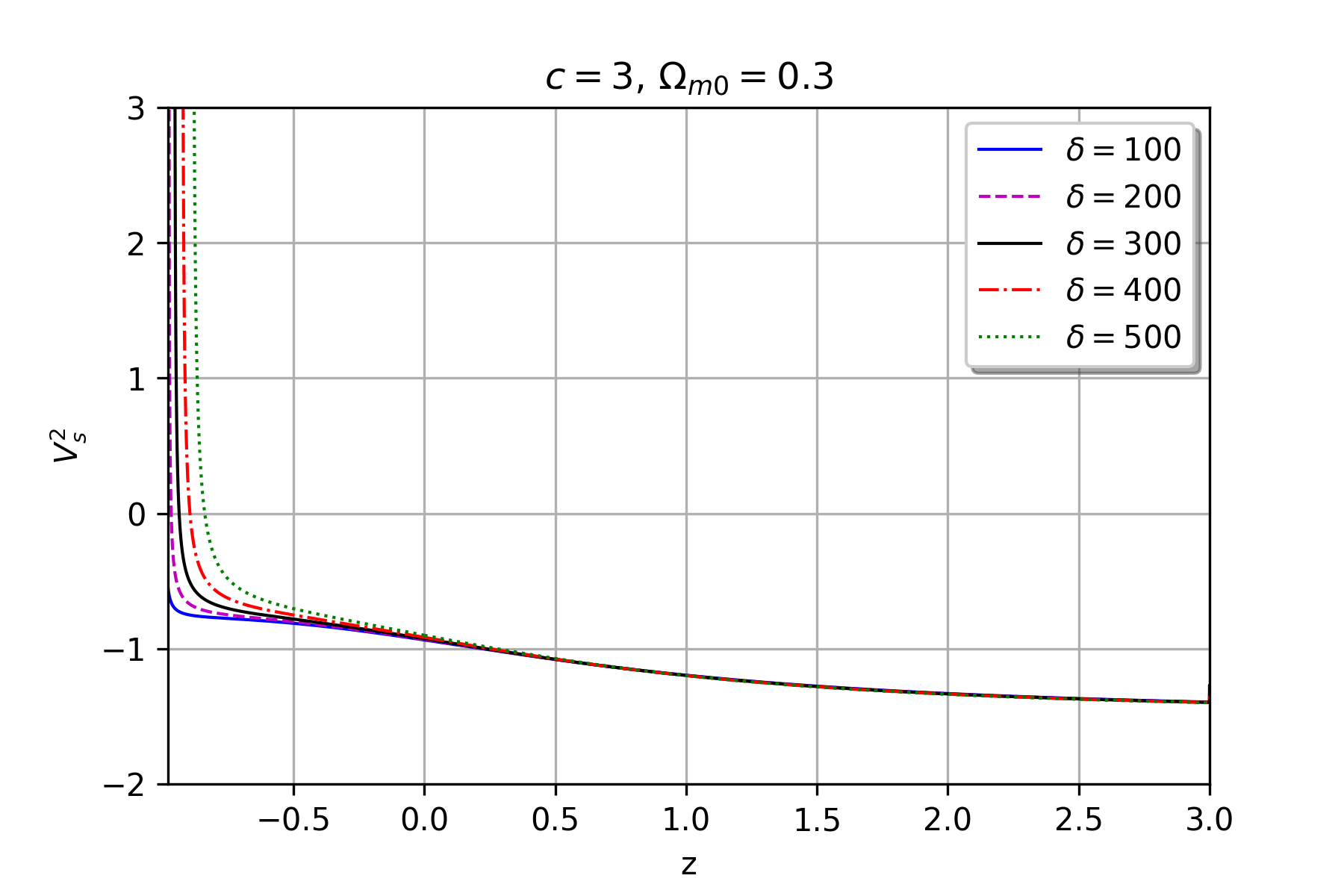}
	
	\caption{\begin{small}
			Behavior of $v_s^2$ vs $z$ for $H(z=0)=67.8, \quad M_p^2=1$ and prescribed values of $K$.
	\end{small}}
	\label{P3.7} 
\end{figure}

\begin{figure}[H]
	\centering
	\includegraphics[width=0.8\linewidth]{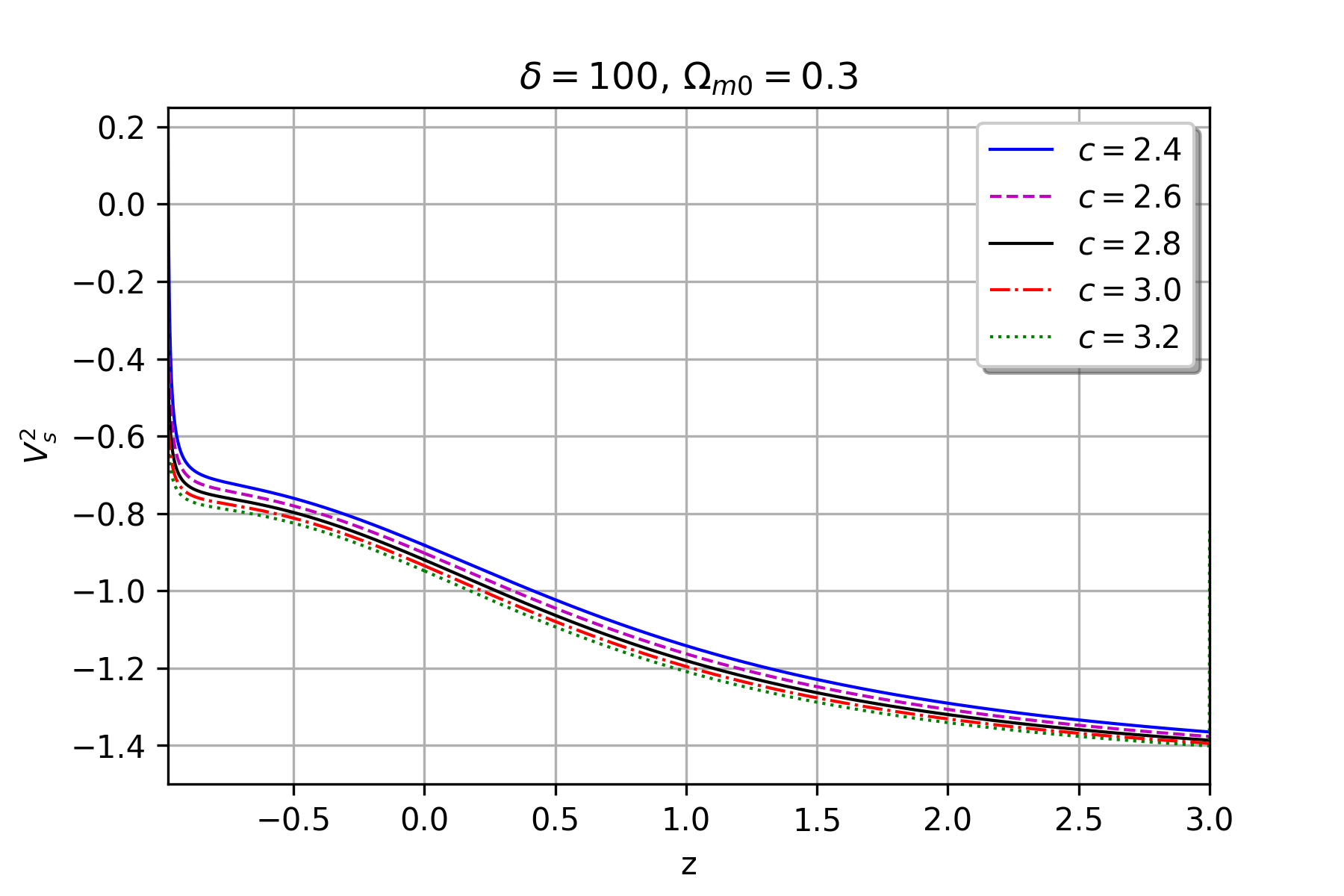}
	
	\caption{\begin{small}
			Behavior of $v_s^2$ vs $z$ for $H(z=0)=67.8, \quad M_p^2=1$ and prescribed values of $c$.
	\end{small}}
	\label{P3.8} 
\end{figure}

The squared sound speed plot in figures \ref{P3.7} and \ref{P3.8} show the non-interacting NTADE model to be unstable for current  cosmic behaviors.

\section{Interacting NTADE Model}
In this section the expressions are formulated by considering a flow of energy among the DM and DE sectors and hence the coupling term $Q$ appears with non-zero magnitude. Here we have chosen $Q=3b^2H(\rho_d+\rho_m)$ \cite{Pavon05,Abdollahi18, Sharma20, Sadeghi13,Honarvaryan15,Goswami:2019zci}.
Differentiating (\ref{3.2}) w.r.t. time and using equation (\ref{3.5}), the expression for the EoS parameter is obtained as 
\begin{eqnarray}
	w_d &&= -1-\dfrac{b^2}{\Omega_d}   \notag \\
	&&-\dfrac{2}{3HT}\left(\dfrac{\delta^2M_p^4T^4-3c^2}{\delta^2M_p^4T^4+\delta M_p^2T^2+3c^2}\right), \label{3.12}
\end{eqnarray}

where $T$ is given by (\ref{3.8a}).\\

Differentiating (\ref{3.6}) w.r.t. time, using (\ref{3.5})and (\ref{3.7}) we get 
\begin{eqnarray}
	\frac{\dot{H}}{H^2} &&=\dfrac{3}{2}\left(-1+b^2+\Omega_d \right)\notag \\
	&&+\dfrac{\Omega_d}{HT}\left(\dfrac{\delta^2M_p^4T^4-3c^2}{\delta^2M_p^4T^4+\delta M_p^2T^2+3c^2}\right), \label{3.13}
\end{eqnarray}
which leads to the deceleration parameter expression as
\begin{eqnarray}
	&& q=\dfrac{1}{2}\left(1-3b^2-3\Omega_d\right)\notag \\ && -\dfrac{\Omega_d}{HT}\left(\dfrac{\delta^2M_p^4T^4-3c^2}{\delta^2M_p^4T^4+\delta M_p^2T^2+3c^2}\right). \label{3.14}
\end{eqnarray}

Taking time derivative of (\ref{3.7a}) and using (\ref{3.5}), the differential equation for the NTADE density parameter is obtained as
\begin{eqnarray}
	&& \dot{\Omega_d}=2H(1+q)\Omega_d \notag \\ &&+\dfrac{2\Omega_d}{T}\left(\dfrac{\delta^2M_p^4T^4-3c^2}{\delta^2M_p^4T^4+\delta M_p^2T^2+3c^2}\right).  \label{3.15}
\end{eqnarray}

\begin{figure}[H]
	\centering
	\includegraphics[width=0.8\linewidth]{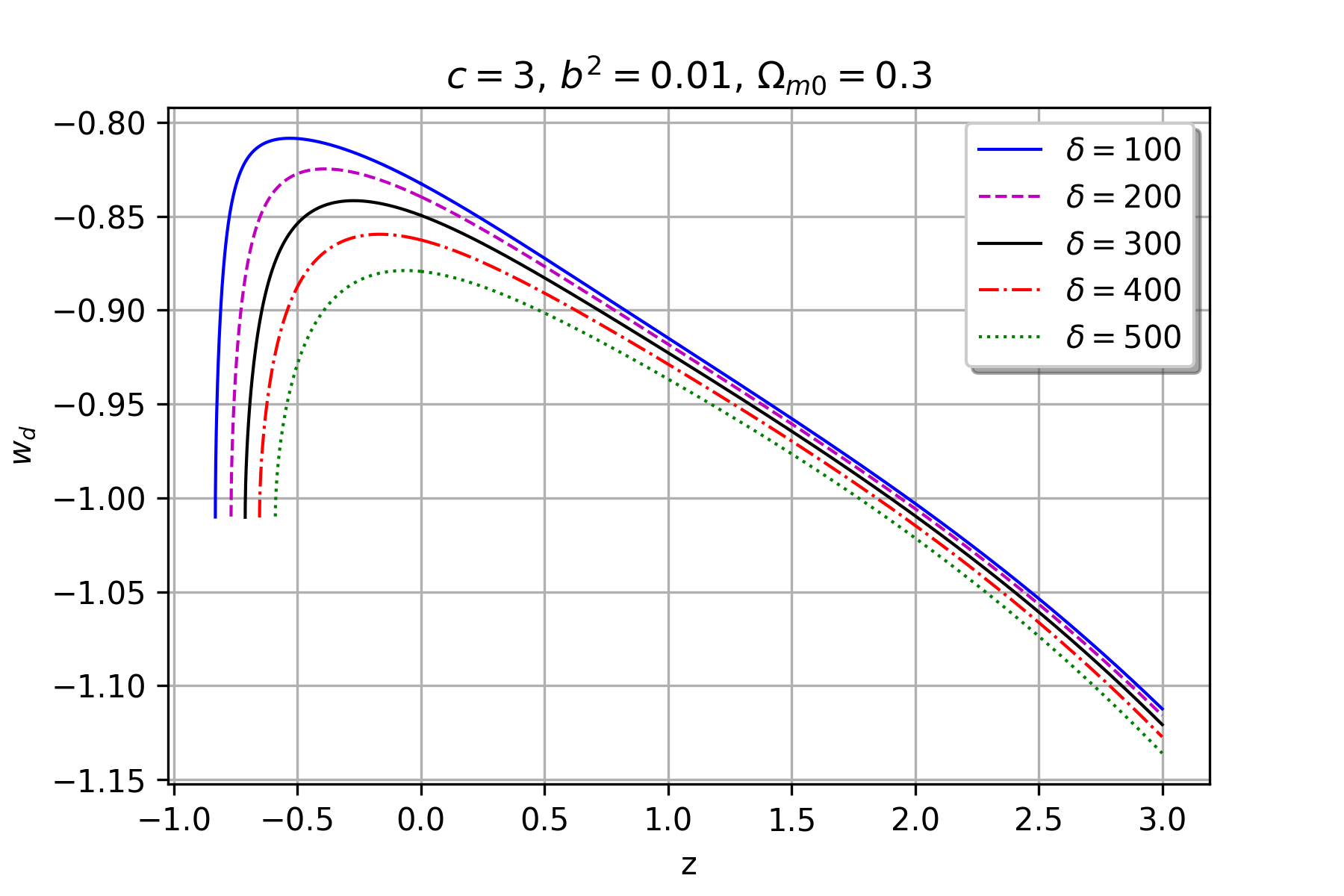}
	
	\caption{\begin{small}
			Behavior of $w_d$ vs $z$ for $ H(z=0)=67.8, \quad M_p^2=1$ and prescribed values of $\delta$.
	\end{small}}
	\label{P3.9} 
\end{figure}

\begin{figure}[H]
	\centering
	\includegraphics[width=\linewidth]{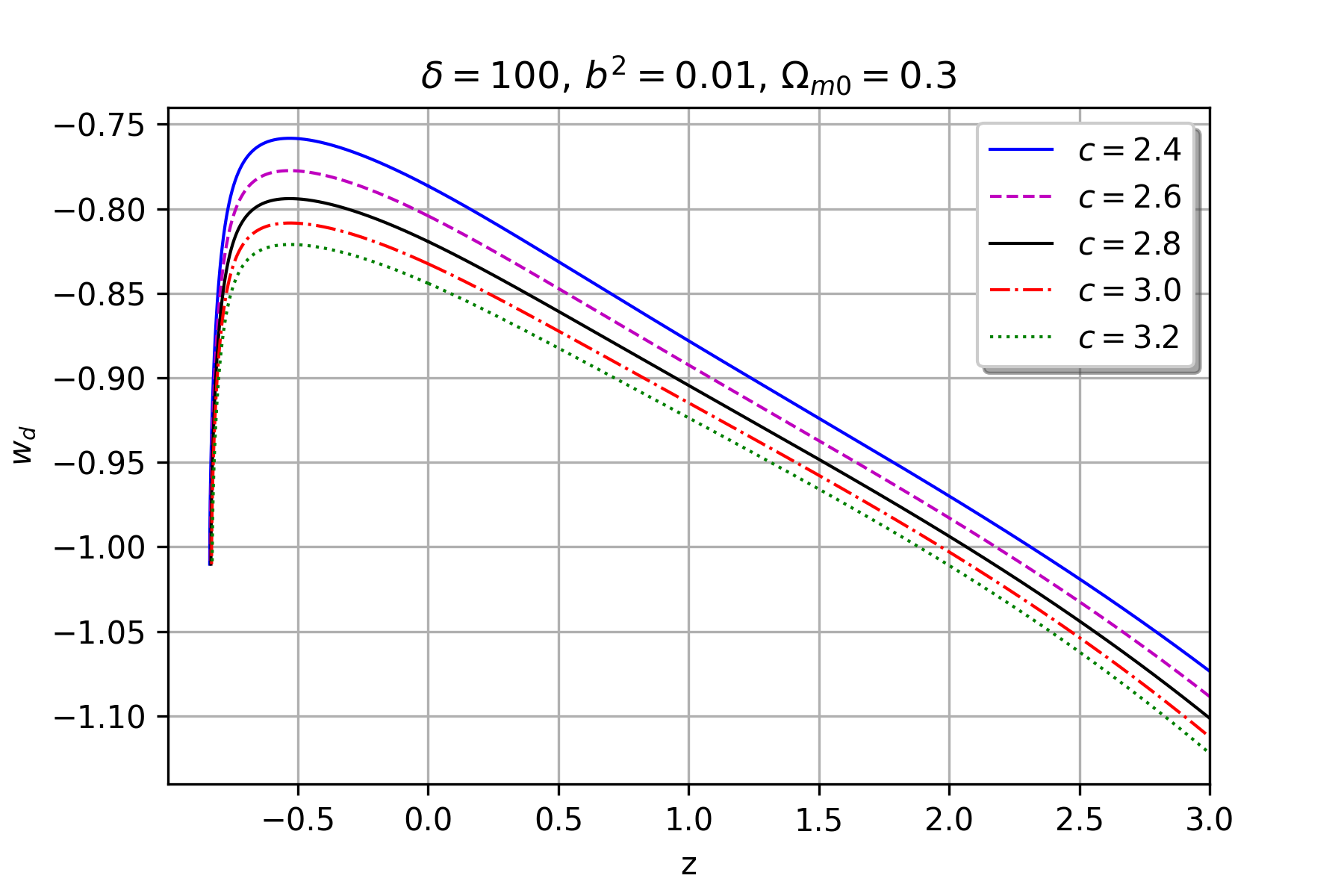}
	
	\caption{\begin{small}
			Behavior of $w_d$ vs $z$ for $ H(z=0)=67.8, \quad M_p^2=1$ and prescribed values of $c$.
	\end{small}}
	\label{P3.10} 
\end{figure}
\vfill
\begin{figure}[H]
	\centering
	\includegraphics[width=\linewidth]{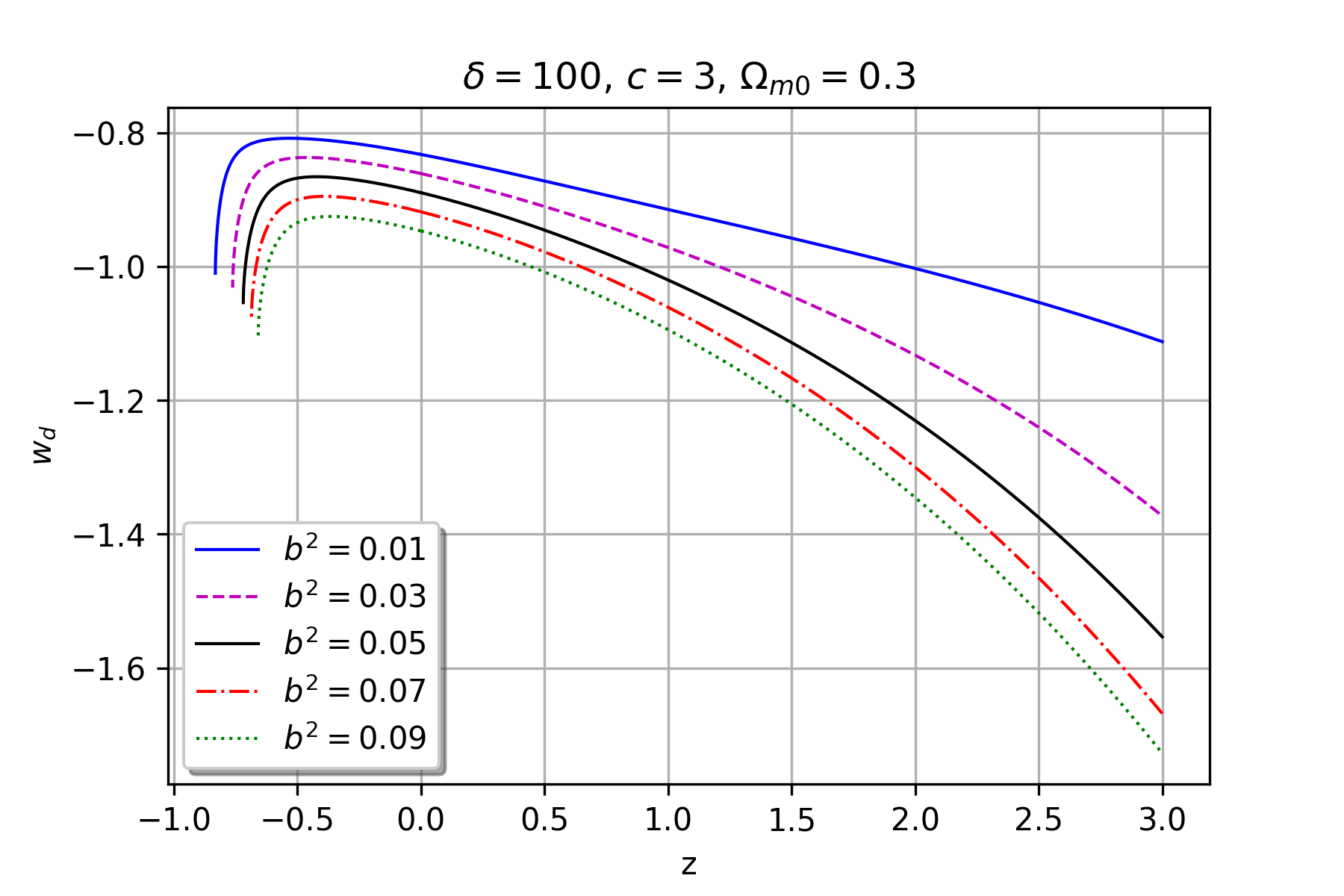}
	\caption{\begin{small}
			Behavior of $w_d$ vs $z$ for $ H(z=0)=67.8, \quad M_p^2=1$ and prescribed values of $b^2$.
	\end{small}}
	\label{P3.11} 
\end{figure}
Behavior of the EoS parameter for interacting NTADE model is shown in figures \ref{P3.9}-\ref{P3.11}. In figure \ref{P3.9}, $\delta$ is varying with fixed $c$ and $b^2$. Similarly, figures \ref{P3.10} and \ref{P3.11} are plotted by varying $c$ and $b^2$, respectively by keeping the other two parameters fixed. Plots in figures \ref{P3.9}-\ref{P3.11} show that the universe evolved in a phantom region.  The present Universe rests in the quintessence zone and intends to cross the divide line $w_d=-1$ in near or far future to enter the phantom zone. In figure \ref{P3.11}, the variation among the curves is more clear than others where $b^2$ is kept constant. This reflects appreciable features of the NTADE model on considering the interaction among the cosmos sectors. 

\begin{figure}[H]
	\centering
	\includegraphics[width=0.9\linewidth]{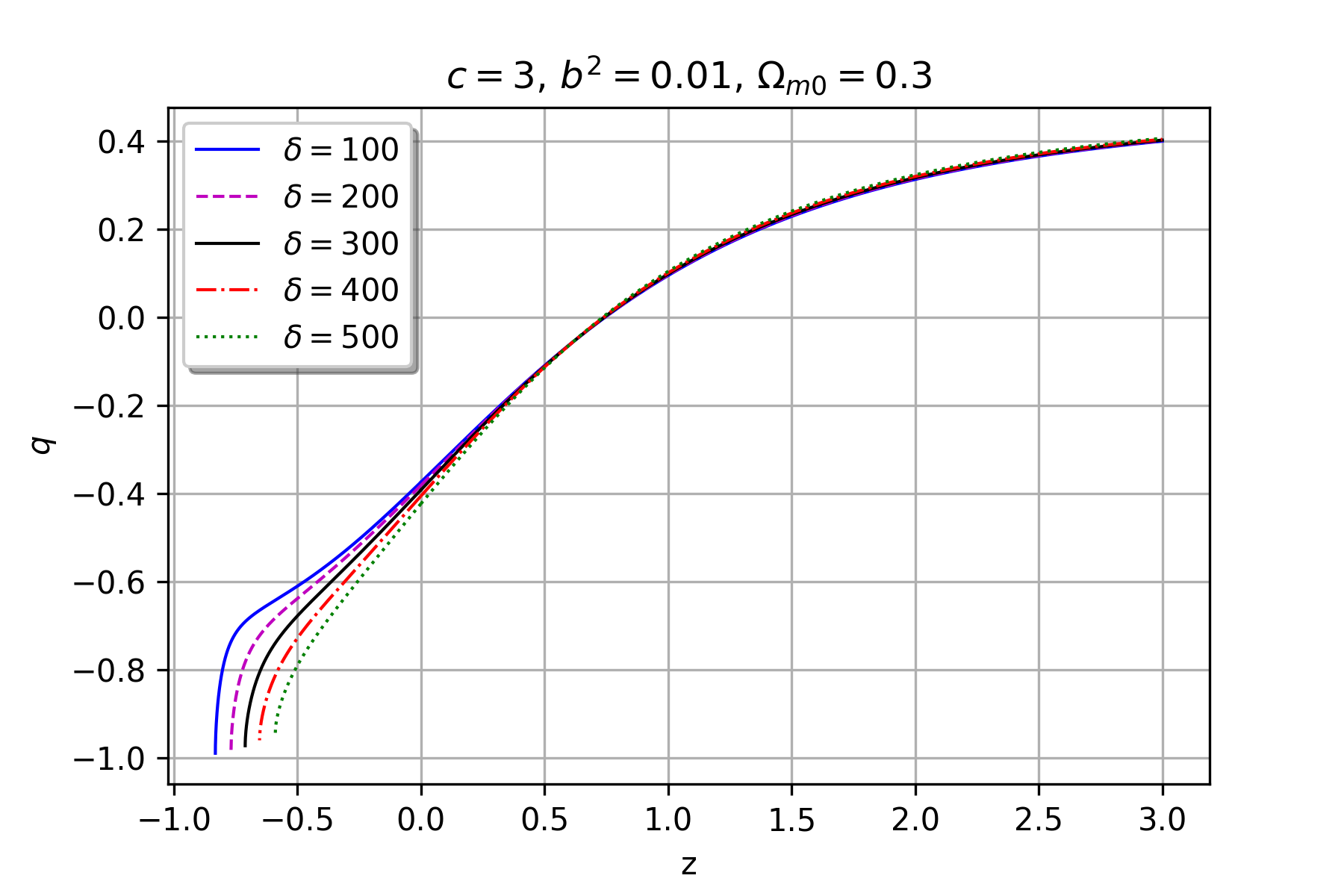}
	
	\caption{\begin{small}
			Behavior of $q$ vs $z$ for $ H(z=0)=67.8, \quad M_p^2=1$ and prescribed values of $\delta$.
	\end{small}}
	\label{P3.12} 
\end{figure}

\begin{figure}[H]
	\centering
	\includegraphics[width=0.8\linewidth]{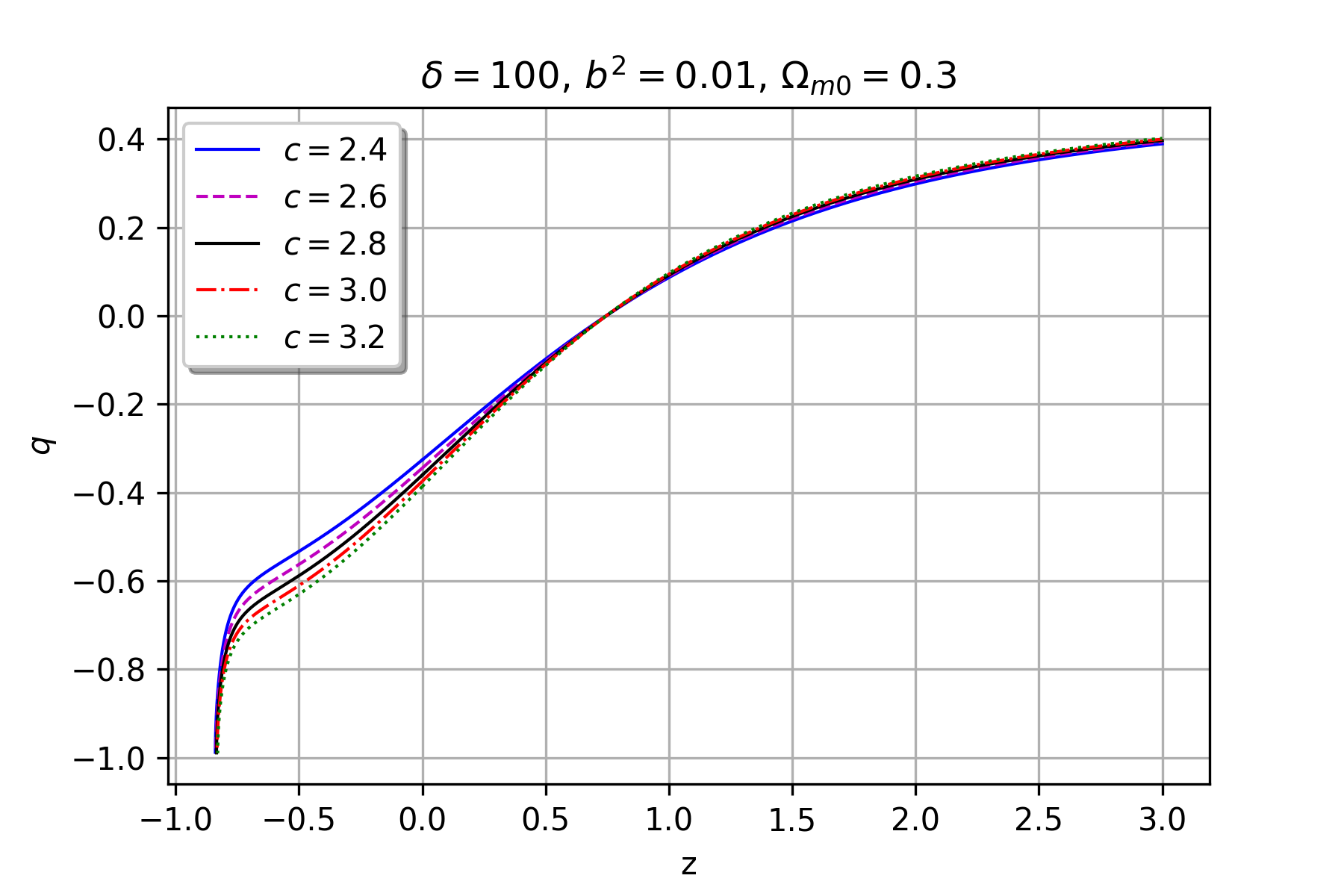}
	
	\caption{\begin{small}
			Behavior of $q$ vs $z$ for $ H(z=0)=67.8, \quad M_p^2=1$ and prescribed values of $c$.
	\end{small}}
	\label{P3.13} 
\end{figure}

\begin{figure}[H]
	\centering
	\includegraphics[width=0.8\linewidth]{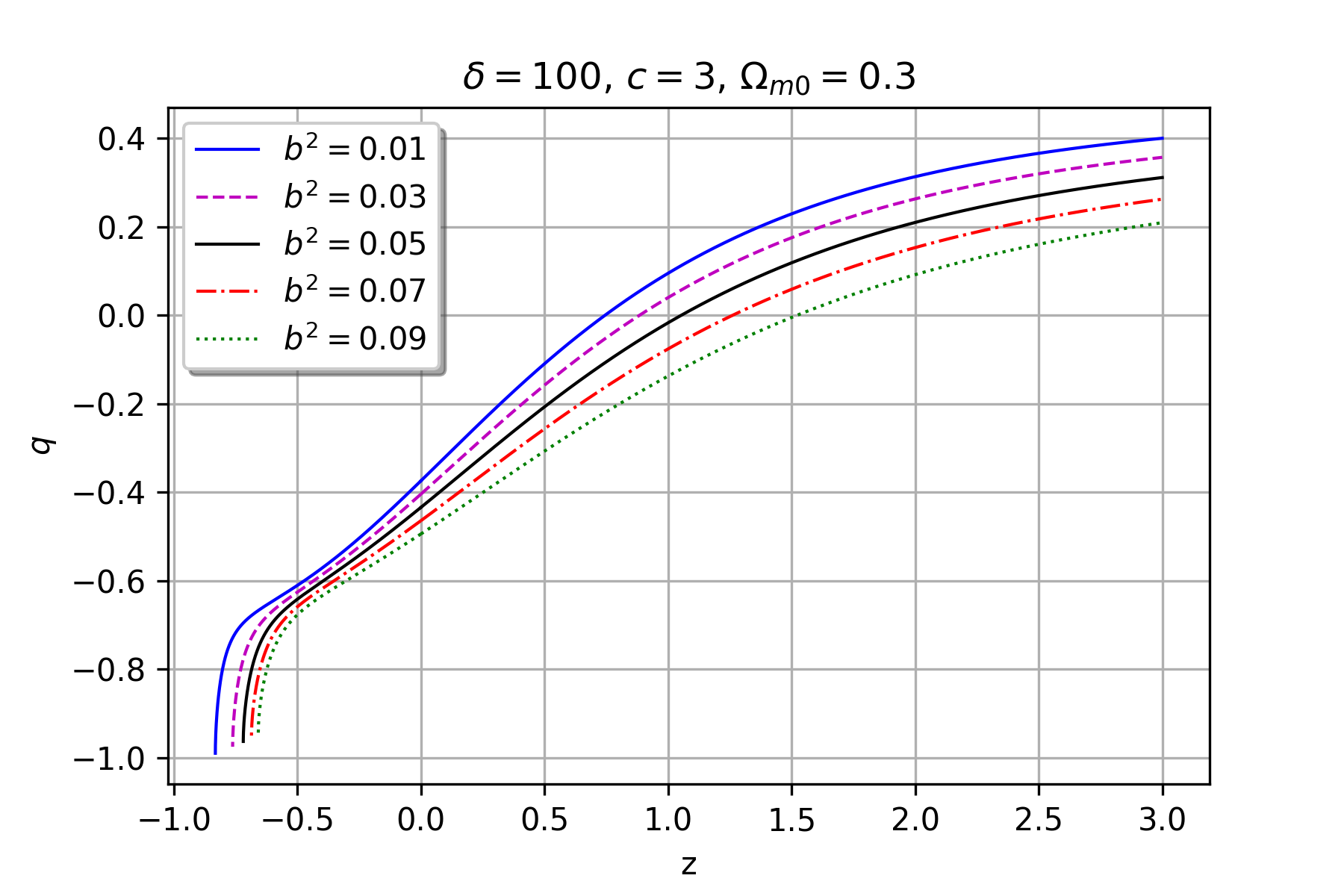}
	
	\caption{\begin{small}
			Behavior of $q$ vs $z$ for $ H(z=0)=67.8, \quad M_p^2=1$ and prescribed values of $b^2$.
	\end{small}}
	\label{P3.14} 
\end{figure}
Figures \ref{P3.12}-\ref{P3.14} are showing the deceleration parameter behavior against the redshift $z$. Plots in \ref{P3.12} and \ref{P3.13} show that in the past, the universe faced the decelerated phase. And, somewhere in $z\in (0.5, 0.8)$, it started accelerating. Both the figures confirm the current stage of the universe to be in accelerated expansion mode. Figure \ref{P3.14} is plotted by varying $b^2$, for fixed $c$ and $\delta$ values. \ref{P3.14} also supports the other two figures in telling the past of the universe to be in the decelerated phase. But, due to the presence of $b^2$ the curves show more variation and hence, the bound for $z\in (0.5, 0.8)$ to enter the universe under accelerated expansion is no longer true, i.e. it widens the transit $z-$ range. But all three figures favor the current and future universe to be in an accelerated phase. As evident, the current $q$ value lies in $z\in(0.2, 0.6)$.

\begin{figure}[H]
	\centering
	\includegraphics[width=0.8\linewidth]{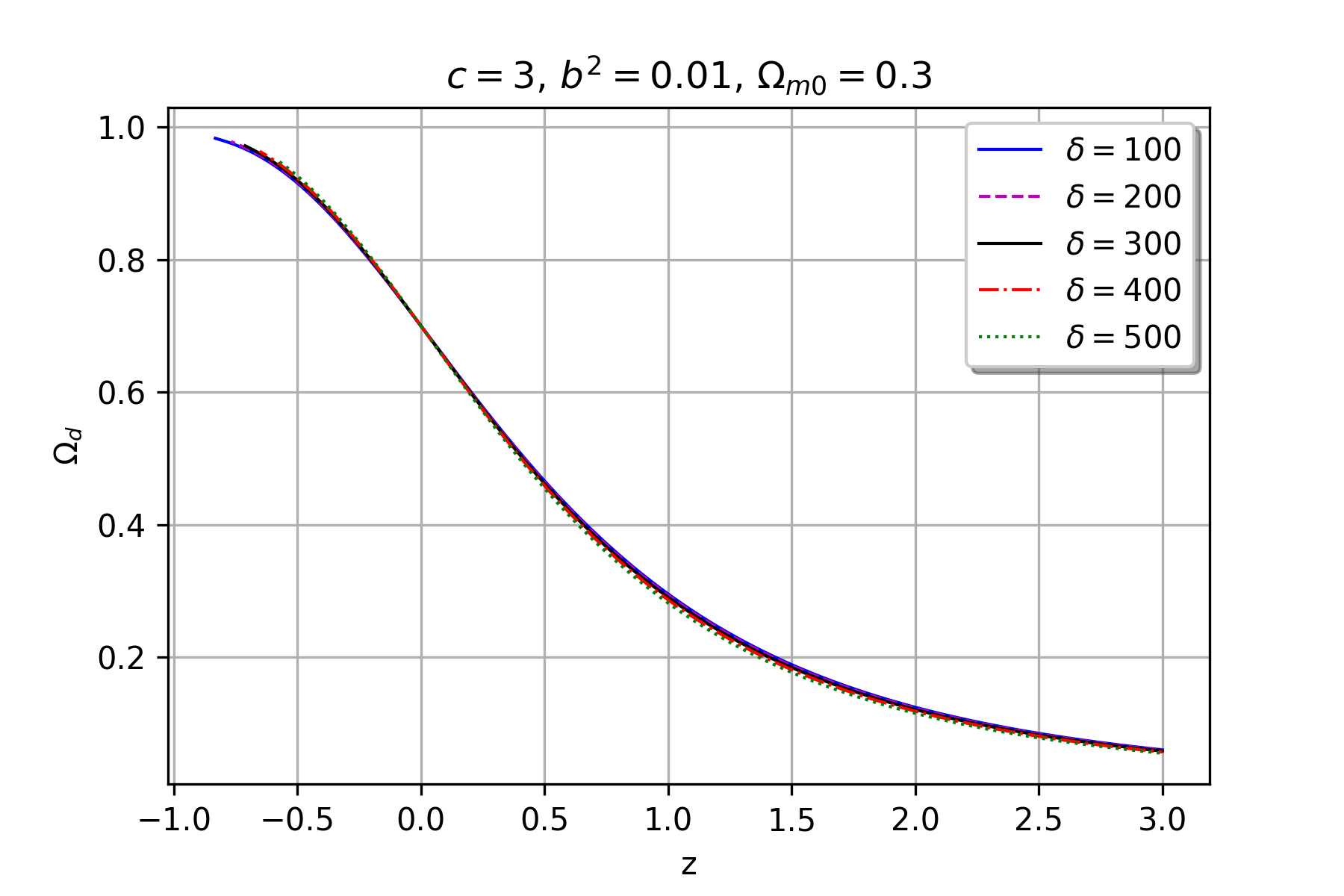}
	
	\caption{\begin{small}
			Behavior of $\Omega_d$ vs $z$ for $  H(z=0)=67.8, \quad M_p^2=1$ and prescribed values of $\delta$.
	\end{small}}
	\label{P3.15} 
\end{figure}

\begin{figure}[H]
	\centering
	\includegraphics[width=0.8\linewidth]{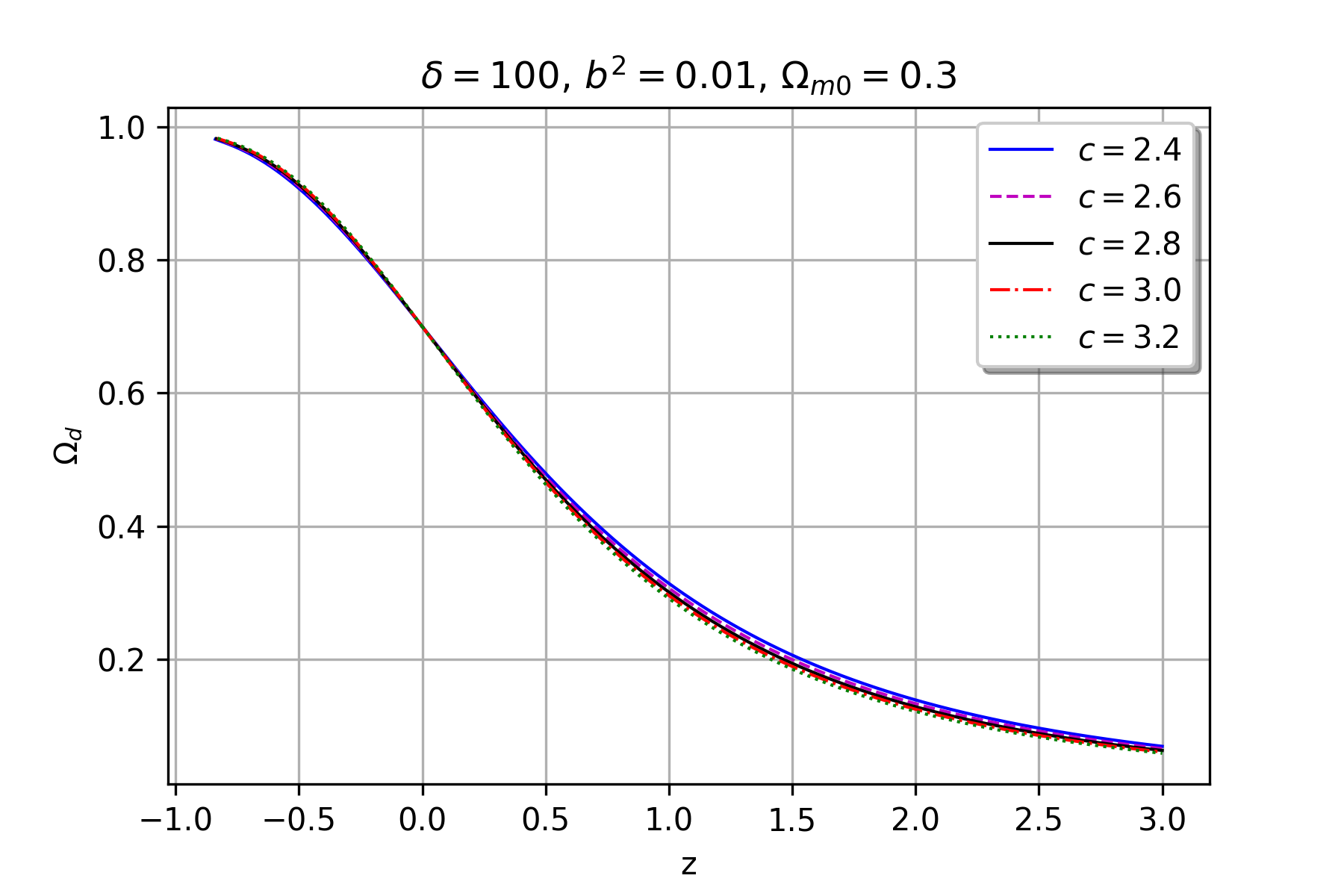}
	
	\caption{\begin{small}
			Behavior of $\Omega_d$ vs $z$ for $ H(z=0)=67.8, \quad M_p^2=1$ and prescribed values of $c$.
	\end{small}}
	\label{P3.16} 
\end{figure}

\begin{figure}[H]
	\centering
	\includegraphics[width=0.8\linewidth]{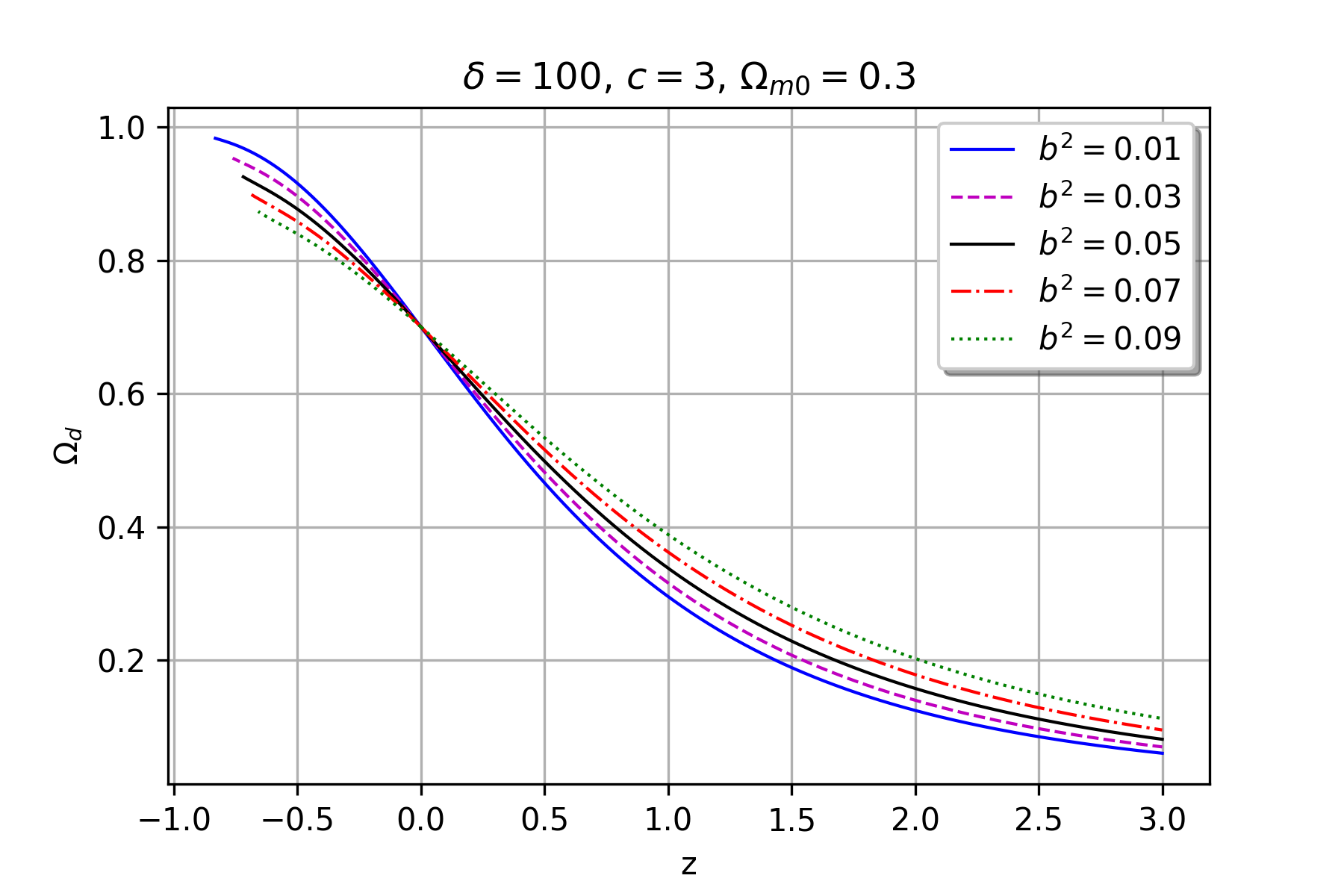}
	
	\caption{\begin{small}
			Behavior of $\Omega_d$ vs $z$ for $H(z=0)=67.8, \quad M_p^2=1$ and prescribed values of $b^2$.
	\end{small}}
	\label{P3.17} 
\end{figure}
The NTADE density parameter $\Omega_d$ explaining the evolutionary behavior of the universe is plotted in figures \ref{P3.15}-\ref{P3.17}. It is clear that the universe in the past was fully dominated by the DM sector. But, by passage of time, NTADE started sharing with gradual increment somewhere in $z \gtrsim 0.5$  \cite{Frieman08}. But the figure with $b^2$ variation shows the NTADE started overtaking the DM share in evolution of the universe somewhere in $z\in(0.2, 0.5)$. All the curves in figures \ref{P3.15}-\ref{P3.17} reflect the current universe to be dominated by its NTADE constituent. They also indicate to occupy the whole share in near or far future by fully occupying the DM sector share.

\begin{figure}[H]
	\centering
	\includegraphics[width=0.8\linewidth]{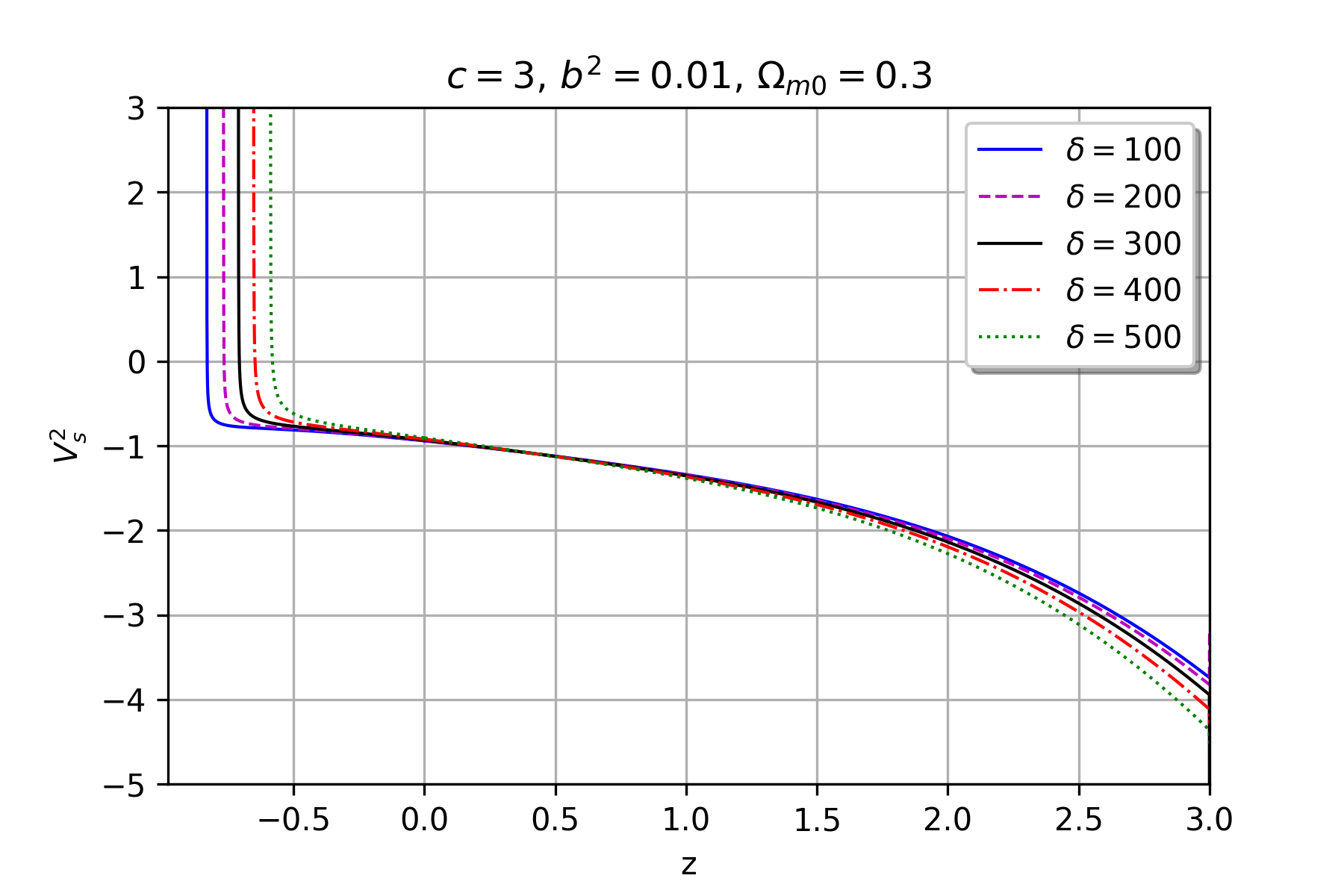}
	
	\caption{\begin{small}
			Behavior of $v_s^2$ vs $z$ for $ H(z=0)=67.8, \quad M_p^2=1$ and prescribed values of $\delta$.
	\end{small}}
	\label{P3.18} 
\end{figure}

\begin{figure}[H]
	\centering
	\includegraphics[width=0.8\linewidth]{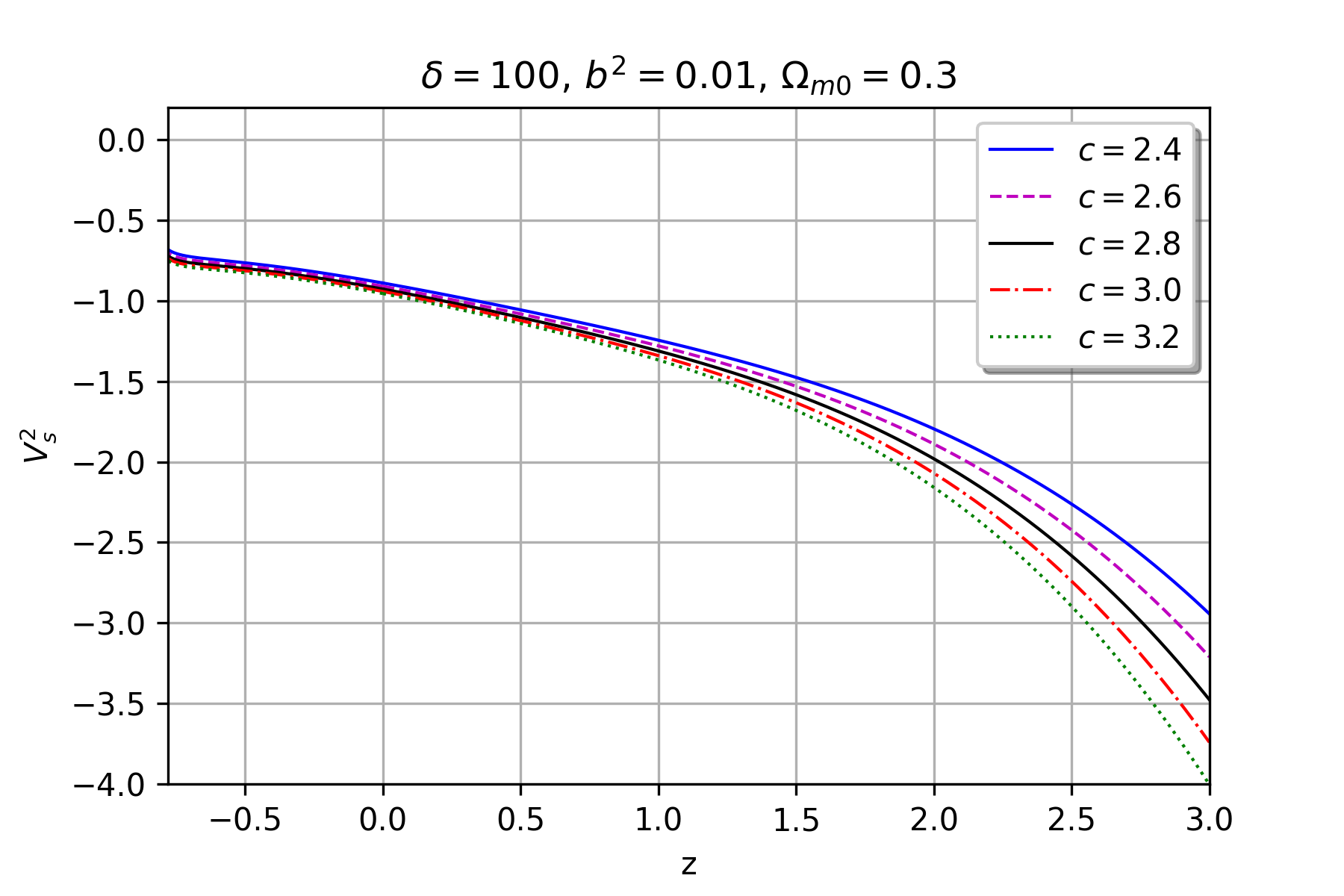}
	
	\caption{\begin{small}
			Behavior of $v_s^2$ vs $z$ for $H(z=0)=67.8, \quad M_p^2=1$ and prescribed values of $c$.
	\end{small}}
	\label{P3.19} 
\end{figure}

\begin{figure}[H]
	\centering
	\includegraphics[width=0.8\linewidth]{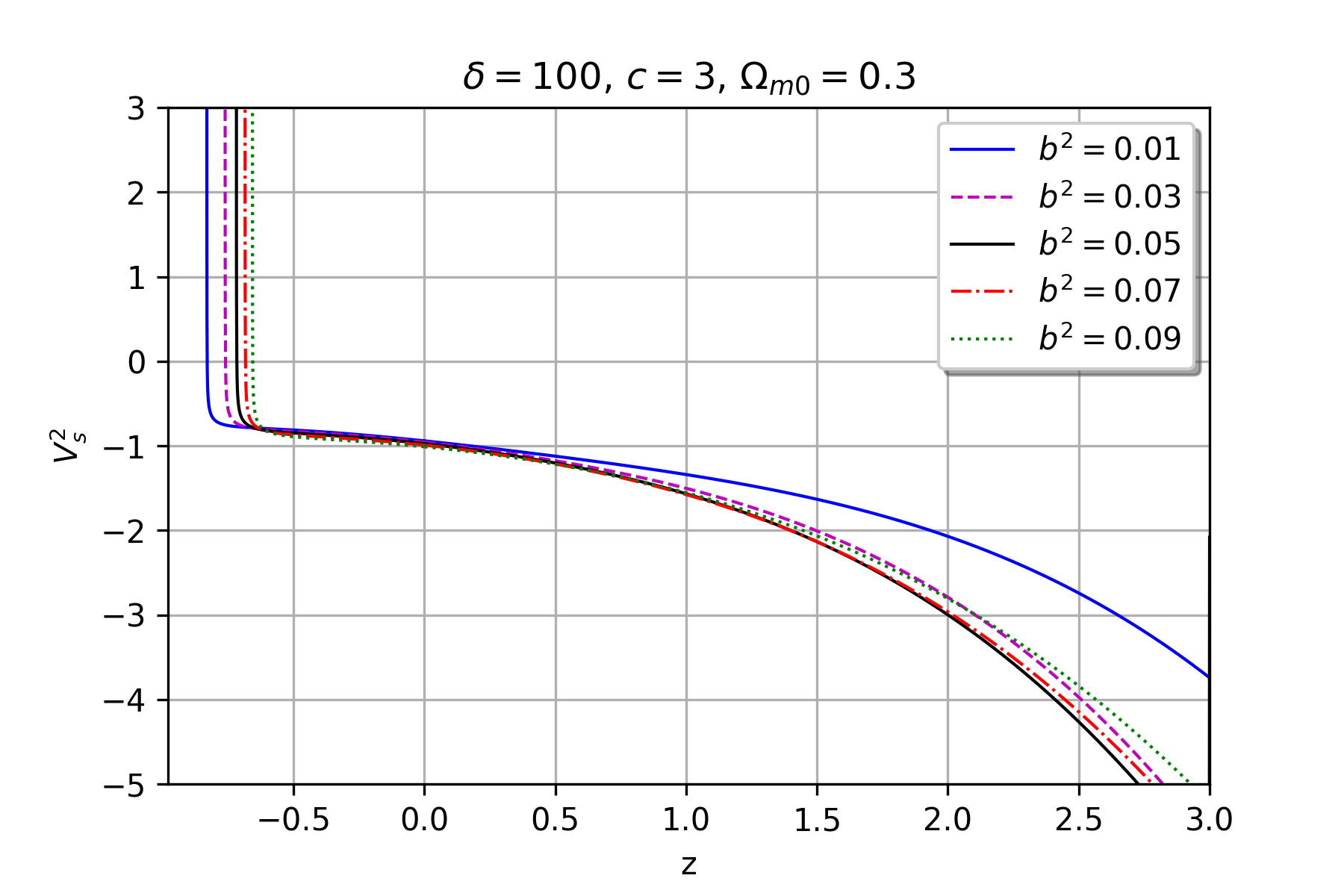}
	
	\caption{\begin{small}
			Behavior of $v_s^2$ vs $z$ for $ H(z=0)=67.8, \quad M_p^2=1$ and prescribed values of $b^2$.
	\end{small}}
	\label{P3.20} 
\end{figure}

The squared sound speed is plotted in figures \ref{P3.18}-\ref{P3.20}. These figures are plotted based on $\delta$, $c$ and $b^2$ variation respectively. These figures show that the NTADE model is efficient for future predictions of the universe.

\section{Summary of the Results}
To analyze the evolutionary behavior of the universe, we proposed the NTADE model as an alternative option to the holographic dark energy models. Being related to long-range interaction behavior of gravity, the non-extensive Tsallis entropy is used. Age of the universe is considered as IR cutoff instead of horizons. The  Karolyhazy uncertainty relation and the original ADE model \cite{Cai07} are the building blocks of the model. The model is studied by considering the independence among  DM and DE sectors constituting the universe as well as an interaction among them. The interaction term $Q$ linking the two sector features are also highlighted. In the case of $Q=0$, the NTADE model behaves like a pure quintessence model as evidenced from figures \ref{P3.1} and \ref{P3.2}. But as the link between the two universe constituents is considered ($Q=3b^2H(\rho_d+\rho_m)$), the behavior of the model becomes completely different. As depicted in figures \ref{P3.9}-\ref{P3.11} the interacting NTADE model behaves like Phantom in the past, quintessence at present and, again, becomes phantom by crossing the divide line $w_d=-1$ in near or far future. The deceleration parameter as depicted in figures \ref{P3.3} and \ref{P3.4} for non interacting cases shows the universe to enter accelerated phase for $z\approx 0.6$. But when interaction is considered, figures \ref{P3.12} and \ref{P3.13} indicate the accelerated expanding phase of the universe to happen somewhere in $z\in(0.2, 1)$. The $b^2$ varying figure \ref{P3.14} shows a broad spectrum for $z$ and hence the interval $z\in(0.2, 1)$ is no longer valid. Which shows further possibilities for the $z-$ range to be considered for the universe to transit into an accelerated phase. The current $q$ value found to be in $(-0.2, -0.5)$ for interacting as well as non interacting scenarios with constant $Q$. As $Q$ starts varying in figure \ref{P3.14}, the current $q$ value shows more variation than the one considered in figures \ref{P3.3}-\ref{P3.4} and \ref{P3.12}-\ref{P3.13}. The NTADE density parameter in figures \ref{P3.5}-\ref{P3.6} and \ref{P3.15}-\ref{P3.16} shows that the past of the universe was dominated by the matter sector $(z \gtrsim 0.5)$. But as $z$ became smaller the NTADE started overtaking the matter sector and reached to $\approx 70\%$ at present and shows the future of the universe to be fully dominated by the energy sector. For $b^2$ variation in figure \ref{P3.17}, scenarios similar to other NTADE density figures are reflected. But, in this case, DM to NTADE takeover in $(z \gtrsim 0.5)$ is no longer valid. Although the Squared sound speed in figures \ref{P3.7}-\ref{P3.8}  shows the instability of the non-interacting NTADE model, as the interaction comes to the picture, the model becomes stable for future $z$ values. And hence we can say, the interaction, i.e. the entry of the parameter $b^2$ increases the efficacy of the model.  The real range of parameters $\delta$, $c$ and $b^2$ are the open problems to be established by observational constraints.

\end{document}